\documentclass[a4paper,11pt]{article}
\usepackage{pos}

\title{Why is Parity \emph{RESTORED}?}
\ShortTitle{Why is Parity RESTORED? }

\author*[a]{Jean-Marie Fr\`{e}re  }

\affiliation[a]{Theoretical Physics CP225, Universit\'{e} Libre de Bruxelles,\\
  blvd du Triomphe , 1050 Brussels, Belgium, \\ and Brout-Englert-Lema\^itre Center, Brussels}

\emailAdd{frere@ulb.be}

\abstract{While Left-Right symmetry (space parity) breaking historically appeared as a surprise, we argue that the real wonder is its restoration in long-distance interactions (at least until we find electric dipole moments!).}

\FullConference{%
  Corfu Summer Institute 2021 "School and Workshops on Elementary Particle Physics and Gravity"\\
  29 August - 9 October 2021\\
  Corfu, Greece
}

\begin{document}
\maketitle
\section{Introduction}
\begin{figure}[h]
\centering
\includegraphics[width=10cm]{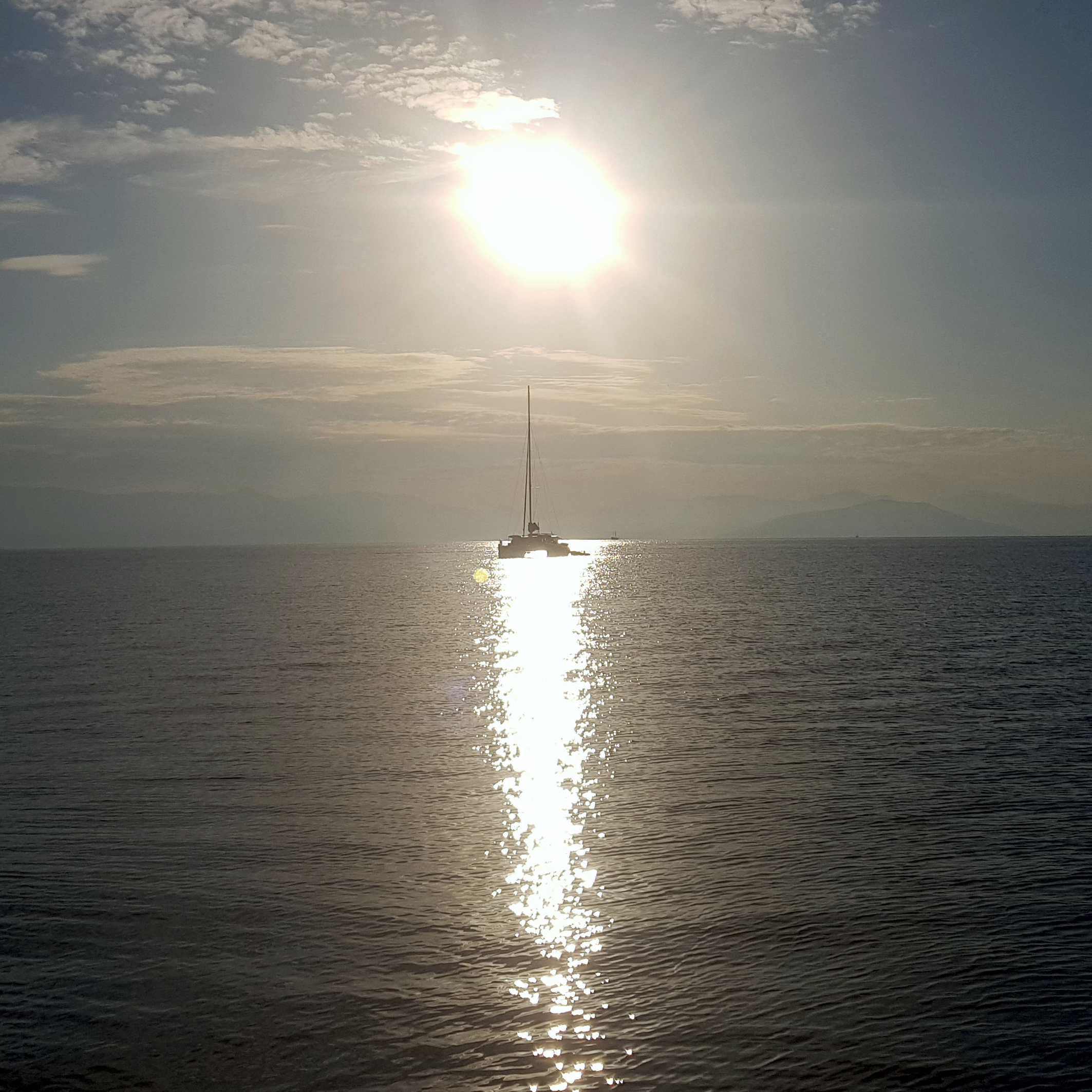}
\caption{Is symmetry in the eye of the beholder?}
\label{boat}
\end{figure}
In a picture taken nearby, the catamaran boat appears symmetrical.
We notice a "natural" tendency to enhance this symmetry, by positioning the camera,
choosing the framing; on the other hand, we tend to ignore the obviously asymmetrical elements (like the clouds),
maybe taking them as "accidental".

Historically, this tendency to search for symmetry has been very strong (I don't allude here to the gauge symmetry
which is not an "apparent" symmetry, but much more a reflection of the redundancy of a representation).
The experimental evidence of symmetry breaking came as a surprise, and efforts were made (for instance: parity doubling) to restore it
even at large cost: we will see other examples.
Interestingly, CP (or T) violation , which may be more bothersome to the mind, was not such a surprise: after the discovery of P violation, it was actively searched for
in  Kaon decays (notably at the suggestion of Lev B. Okun).

While we may tend to believe that delving deeper to more fundamental aspects would reveal more symmetry, in a way, daily practice may teach the opposite.
For instance, humans and most animals exhibit an external (approximate) symmetry, but the internal organs are strongly asymmetrical (not to mention the DNA molecule).
Evolution and natural selection may of course explain that advantages of this symmetry (for locomotion for instance) have imposed the external morphology.

To some extent, this is a metaphor of what happens in (particle) physics: long distance interactions like gravity, electromagnetism and even the strong force act in a symmetrical
way, which may have given the above-mentioned intuition of a fundamental character of parity, while the short-distance interactions (weak interactions) are maximally parity-violating.

While some popular lore tends to blame Left-Right symmetry breaking on some "environmental" circumstance (the de facto absence of a light right-handed neutrino),
the origin is obviously deeper (as we remind that parity breaking was first encountered in purely hadronic interactions), but could still be cured at higher energies
in thus-far elusive L-R symmetrical models (not to mention SO(10) and related approaches).

In an interesting twist, we might \emph{a contrario} find some residual fairly long-range parity violation through electric dipole moments for instance.
\section{It came as a surprise ...and in the hadronic sector}
Although some early indications of polarization of beta rays from radium decays existed,
evidence from the purely hadronic sector indicating possible parity violation were a shock.
Known as the $\Theta -\tau$ puzzle, (nothing to do with the $\tau$ lepton) the observation seemed to indicate
the presence of 2 distinct  spin-0 particles produced under the same conditions, of similar mass but decaying differently: in 2 or 3 pions (in s-wave).
$$\Theta^+ \rightarrow \pi^+ \pi^0  ; \tau^+ \rightarrow \pi^+ \pi^+ \pi^- $$

Since the pion parity had been fixed to (-1) from its interactions with nucleons, this implied either 2 degenerate particles (up to the experimental precision) or a parity violation, as proposed by T.D. Lee and C.N.Yang.

The textbook C-S. Wu experiment then came as a test and a confirmation of the 2nd interpretation, rather than a discovery per se.
\begin{figure}[h]
\centering
\includegraphics[width=4cm]{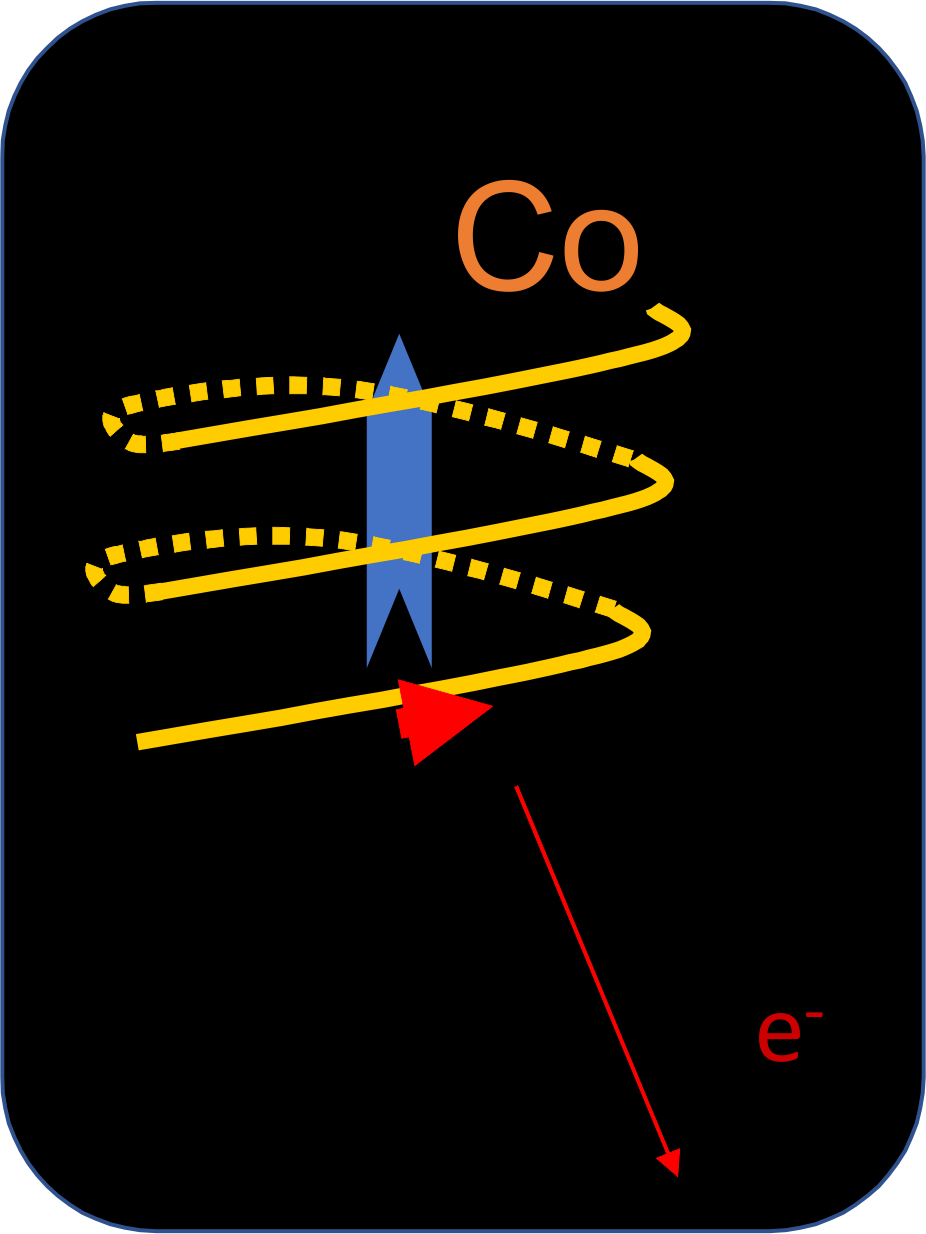}
\caption{Schematics of the Wu experiment}
\label{Wu}
\end{figure}

The flux of electrons is maximum  opposite to the decaying Co spin, which ensures that the average of the pseudoscalar observable $<p_e \bullet s_{Co}>$
(which should vanish if parity were conserved) is non-zero.

\begin{figure}[h]
\centering
\includegraphics[width=6cm]{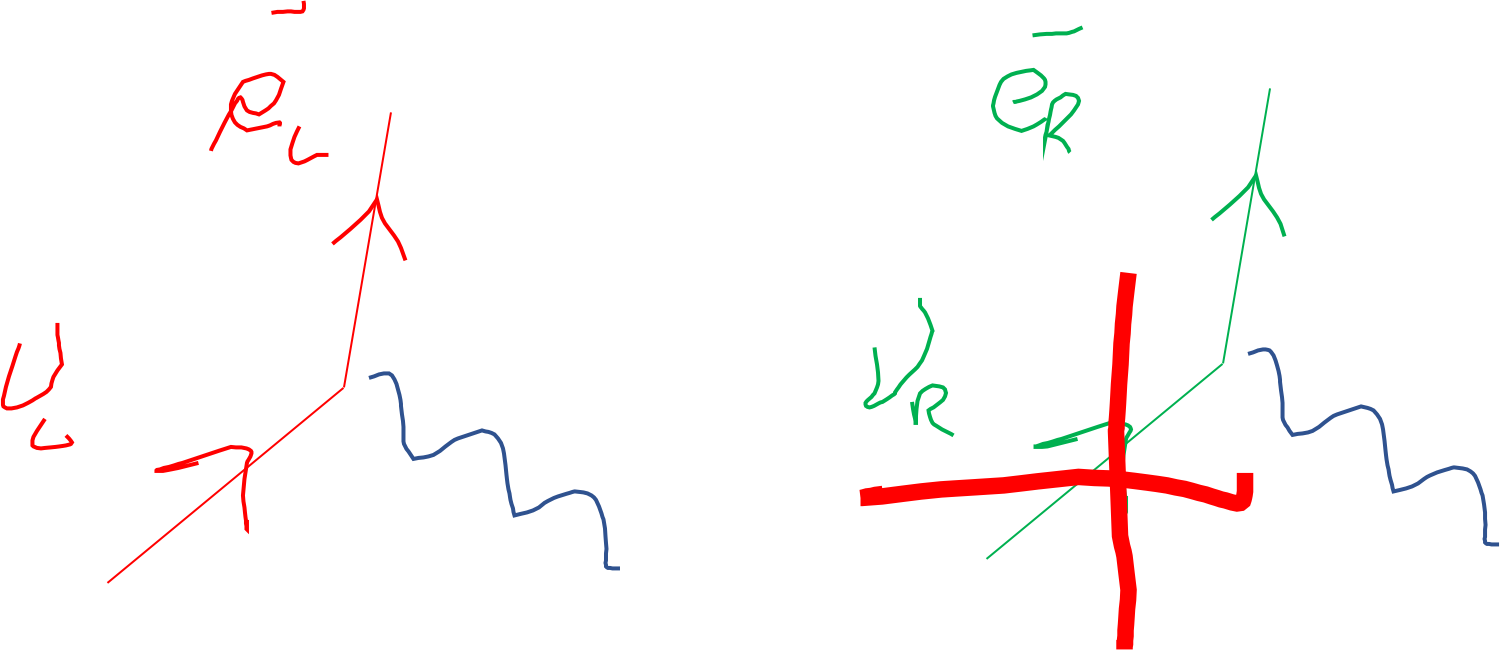}
\caption{Should we blame the neutrino?}
\label{MissingNR}
\end{figure}
A straightforward interpretation is in the nature of the weak boson W couplings.
This can however be formulated in 2 slightly different but fundamentally inequivalent ways. 
A popular saying was that the W boson couples in a L-R symmetrical way (Vector coupling) but that the 2nd process is
forbidden by say, a large Majorana mass for the $\nu_R$. This is a typical example of the approach "fundamental laws
have to be symmetrical but accidental -boundary- conditions may not". (this even applies to the expanding universe!) 
Alternatively but still in the same line of reasoning the $\nu_R$ might even be absent, although this attitude in fact
implies the 2nd approach. Very often it is still maintained (even in teaching) that the $\nu_R$ is "beyond the Standard Model" for this reason.
If it is true that it was not included in the original papers, neither were the heavy quarks: they were either not known or not needed. 
It can certainly be argued that having $\nu_R$ and Dirac masses for neutrinos present is the simplest way to account for massive neutrinos, 
and the one most in line with the Standard Model, even though Majorana masses are an elegant possibility. 

\section{Don't blame the neutrino}

The 2nd approach (which proves the correct one) is to assume that, irrespective of $\nu_R$'s presence, the W bosons only couple
to the left-handed leptons. By this, we imply that not only $W^\pm$ but also (by gauge invariance) their SU(2) partner $W^0$, which makes part of the photon and the neutral Z boson, has purely chiral couplings to leptons. 
This approach was quickly vindicated by the discovery of atomic parity violation, which results from the photon-Z interference.(no neutrino
appears in this process).

It also opens the way to explaining the $\Theta -\tau$ puzzle, with both now identified to the $K^\pm$ , and the chiral couplings of 
the quarks explaining this purely hadronic decay parity violation (which would be impossible to account for from the sole absence of $\nu_R$) 

In  passing, I would like to make a remark about "maximally violated parity symmetry" (which is the case in charged weak interactions), to stress that this is in fact to be expected. Since we are dealing with a non-abelian gauge symmetry, the coupling constant is uniquely fixed 
(up to trivial Clebsch-Gordan coefficients). There is no way L and R fermions can couple in slightly different ways, it is an all-or-nothing situation (and, of course, L and R spinors are our fundamental building blocks, not the resulting V or A couplings. Any "intermediary" level of parity violation could only stem from the U(1) sector, but this is strongly constrained by anomalies. 

\section{So...why is Parity \emph{restored}at long distances?}

We have seen above that Parity violation is (for now) a fundamental characteristic of fundamental interactions, and not 
an "accidental", "environmental" or "boundary condition" issue (like the expanding Universe probably is). 

For instance  unified theories, like SU(5), (which would need extra particles to reach grand unification at an acceptable scale) 
is formulated using only left-handed multiplets (which can include CP conjugate spinors). 

Despite this, it is still tempting to advocate a fundamental L-R symmetry which would be broken down to our existing observations.
Such is the case of SO(10)-based models, where the \textbf{16} representation includes all fermions from one family, including $\nu_R$, 
and which restores the symmetry at Lagrangian level. In this case, the group structure results in another kind of "parity doubling". 
Instead of doubling the "matter" objects, we are doubling the "interaction vectors" , allowing for instance for $W_L$ and $W_R$ and
blaming the observed parity violation not on the Lagrangian, but on a peculiar symmetry breaking solution which yields larger masses for the latter boson. This possibility is certainly much alive, despite the problems its implementation may pose with domain walls in the evolution of the early universe.

Now, let us turn to the question announced in this section: \textbf{"how comes we have been abused for so long in thinking that fundamental interactions were L-R symmetrical (parity invariant)?"}

The fact is that all observed interactions, until the discovery of radioactivity (and $\beta$ decays in particular) respected parity. 
This is certainly the case of gravitation, but also of electromagnetism (here the use of a handedness choice to define the B field
is purely conventional, since the same convention comes again in the Lorentz force, thus making sure that observable effects -at the 
difference of man-made tools - are parity invariant). 
Another (almost) macroscopic force, the strong force, which is of relatively long range, also respects parity: \textbf{all was thus put in
place to dupe us into believing that parity invariance was a fundamental requirement of nature}, and the observed asymmetries (DNA, 
asymmetrical internal organization of human beings, ...) were purely accidental (as, for instance, the rotation of the Earth). 

For the Strong force, obeying the gauge symmetry SU(3), a first "explanation" might come from the compensation of anomalies. As long as we only face low representations, the only compensation of the triplet $3$ of left-handed colored quarks occurs through a left-handed $\overline{3}$, which is also the CP conjugate of the right-handed $3$ : this results in a LR symmetrical matter content. 

Probably more relevant is the fact that the presence of mass for the "matter" particles implies LR symmetry. 

Consider indeed the mass term and the gauge transformations (written in a generic way): 
\begin{eqnarray}
   m \ \overline{\psi_L} \psi_R + h.c.\\
  \psi_L \rightarrow e^{i \alpha_L}\psi_L  \\
   \psi_R \rightarrow e^{i \alpha_R}\psi_R \\
\nonumber \Rightarrow \alpha_L = \alpha_R
\end{eqnarray}

For abelian theories (for instance, QED, whether it originates from an abelian formulation, or is left over after breaking of a larger group), this simply implies the equality of L and R couplings, and thus the restoration of parity.
If we try to apply this to non-abelian groups (think SU(3)), since we are in an all-or-nothing situation (the gauge coupling is unique), this 
means that the L and R partners must belong to identical representations. 

Of course, this approach is rather pragmatic: all observed electrically charged particles are massive (the issue is more difficult to decide for quarks, since the determination of masses comes mostly from current algebra, but a massless u quark is generally considered as impossible, and anyway anomaly cancelation in the Standard Model would imply the same result). The argument in any case would apply to the effectively observed objects (say, nucleons). 

For the more speculative minds, this begs the question: is it possible to have massless charged particles? 
We know that such theories are difficult to formulate in a consistent way. Longitudinal divergences are a major concern when no low-energy cut-off is present. For instance, in formulating a scattering experiment, care should be taken that an initial "massless $e_L$" would 
be degenerate with a "massless $e_R$ + L-polarized photon". 

For the time being (but weakly coupled sectors like dark photons might bring surprises) Nature seems to have spared us this chore!

\section{A remaining possibility for long-distance P violation}

One possibility still exists to find long-distance (or at least macroscopic) P violation. 
If indeed the currently searched-for electric dipole moments are found (for the neutron, electron or proton), they would 
induce parity-violating observations, as exemplified in the "gedanken" experiment below. We leave it to the reader to
apply a mirror transformation to the apparatus and compare to the mirror image of the drawing. 

\begin{figure}[h]
\centering
\includegraphics[width=10cm]{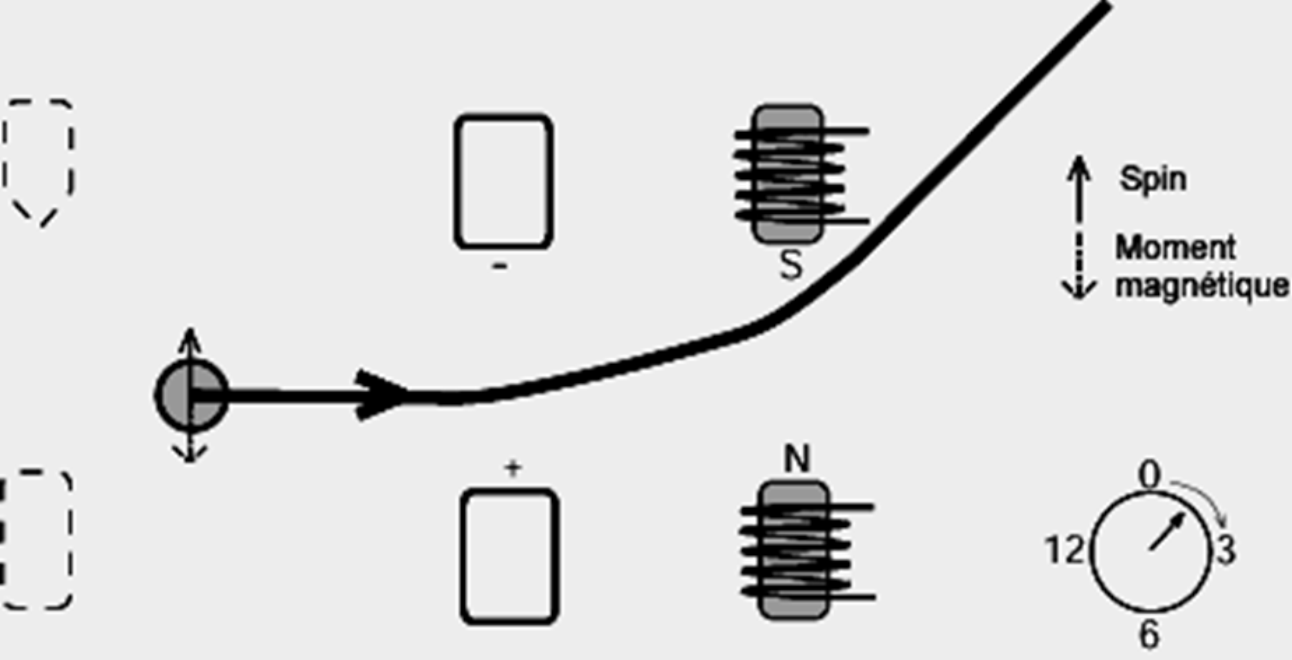}
\caption{Gedanken experiment with neutron dipole moment}
\label{dipole}
\end{figure}

In the present context of course, we would not consider this as a \emph{fundamental interaction}, but rather as a modification
of the electromagnetic sources originating from a (parity violating, possibly spontaneously broken) theory. 

\section{Acknowledgements}
This work is supported by IISN (Belgium) , the Brout-Englert-Lema\^{\i}tre Center (Brussels) funded in part by Innoviris.
I wish to thank Cesar Gomez (uam) for discussions and the organizers of the Corfu meetings for the occasion to evoke those issues.



\end{document}